\documentclass{iau} 
\usepackage{graphicx}
\usepackage{sidecap}
\usepackage{wrapfig}

\title[Rediscovering the Galactic outer disk with LAMOST data] 
{Rediscovering the Galactic outer disk with LAMOST data}

\author[C. Liu et al.]   
{Chao Liu$^1$, Yan Xu$^1$,
  Haifeng Wang$^{1,2}$ \and Junchen Wan$^{1}$}

\affiliation{$^1$Key Lab of Optical Astronomy, National Astronomical Observatories, CAS \\ 20A Datun Road, 100012, Beijing, China \\ email: {\tt liuchao@nao.cas.cn} \\[\affilskip]
$^2$University of Chinese Academy of Sciences, \\ 100049, Beijing, China}

\pubyear{2017}
\volume{334}  
\jname{Rediscovering our Galaxy}
\editors{C. Chiappini, I. Minchev, E. Starkenberg, M. Valentini, eds.}
\begin{document}

\maketitle

\begin{abstract}
From the derived stellar density profile using LAMOST giant stars, we find that the Galactic disk does not show truncation or break, but smoothly transit to the halo from 19 kpc. The scale length of the outer disk is only $1.6\pm0.1$\,kpc, substantially smaller than previous results. This implies that the shapes of the inner and outer disk are different. Meanwhile, the disk flaring is not only found in older populations, but also in younger population. Moreover, the vertical oscillations of the disk are identified in a wide range or $R$ from 8 to 14 kpc. We also find that the velocity dispersion profile as a function of the Galactocentric radius is flat with scale length of $26.3\pm3.2$\,kpc. We confirm that the radial velocity profile in outer disk is significantly affected by asymmetric motion. The bar with either a slower or a faster pattern speed can induce the similar radial asymmetric motion.
\end{abstract}

\firstsection 
\section{Introduction}\label{sec:intro}

\begin{wrapfigure}[20]{r}[0pt]{0.4\textwidth}
\vspace{0pt}
	\centering
	\includegraphics[scale=0.4]{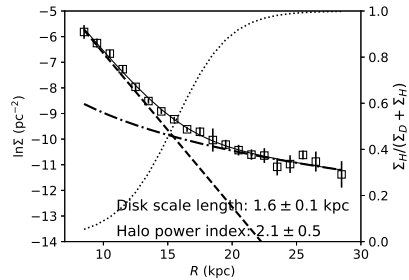}
	\caption{The surface stellar density profile is displayed with the hollow rectangles. The dashed and dot-dashed lines represent for the best fit exponential disk and power-law halo models. The black solid line corresponding to the right-side y-axis stands for the fraction of the stellar halo at given radius (also see \cite[Liu et al. 2017b]{liu2017b}).}\label{fig:1}
	\vspace{15pt}
\end{wrapfigure}

The Galactic disk can be observed in various aspects. One way to shape the disk is using star counting. The vertical density profile can be fitted with a power-law halo profile and two exponential profiles (or ${\rm sech}^2$ profile, see \cite[van der Kruit 1988]{kruit1988}), corresponding to the thin and thick disks (e.g. \cite[Gilmore \& Reid 1983]{gilmore1983}, \cite[Juri\'c et al. 2008]{juric2008}). With more accurate star counting in the solar neighborhood, \cite[Widrow et al. (2012)]{widrow2012} found that the residual of the vertical stellar density profile after subtracting the best-fit model shows clear oscillations.
\cite[Xu et al. (2015)]{xu2015} found that the outer disk beyond the solar neighborhood is also in oscillation. They discovered that the star count in the north of the Galactic mid-plane is more than  that in the south at 2-3\,kpc beyond the location of the Sun (north near substructure) and at the distance of around 10\,kpc (which is the well known ``Monoceros ring'', also see \cite[Newberg et al. 2002]{newberg2002}). Meanwhile, the star count in the south become larger than that of north at the distance of about 5 kpc (south middle substructure).


Another way to understand the nature of the disk is through stellar kinematics. Similar to the stellar density profiles, the velocity distribution in the solar neighborhood is not smooth but clumpy. One of the most prominent velocity substructure is the Hercules stream (\cite[Dehnen 1998]{dehnen1998}), which is believed to be associated with the central bar (\cite[Dehnen 2000]{dehnen2000}, \cite[P\'erez-Villegas et al. 2017 etc.]{perez2017}). However, different studies derived different pattern speeds by interpreting the Hercules stream in different way. Not only in the solar neighborhood, \cite[Liu et al. (2012)]{liu2012} also found that the radial velocity distribution is bifurcate at Galactocentric radius of 10.5-11\,kpc, hinting that the perturbation may also occur in a larger range of the disk.

\begin{wrapfigure}[17]{r}[0pt]{0.4\textwidth}
	\centering
	\includegraphics[scale=0.4]{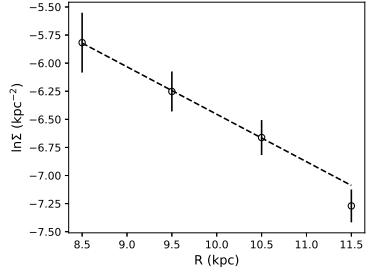}
	\caption{The zoom-in surface stellar density profile at $R<12$\,kpc is displayed with the hollow circles. The dashed line is the best fit exponential model with scale length of $2.37\pm0.02$\,kpc using the first three points.}\label{fig:2}
\end{wrapfigure}

If the disk is axisymmetric, then one would expect that the mean radial and vertical velocities should be zero at any point. However, \cite[Siebert et al. (2011)]{siebert2011}, \cite[Carlin et al. (2013)]{carlin2013}, and \cite[Williams et al. (2013)]{williams2013} found both mean radial and vertical velocities shift from zero in various way at different radii and heights above/below the mid-plane within about 1 kpc around the Sun. These asymmetric motions are probably due to the perturbation of the minor merger (\cite[G\'omez et al. 2013]{gomez2013}),  bar (\cite[Grand et al. 2015]{grand2015}), and/or the spiral structure (\cite[Faure et al. 2014]{faure2014}, \cite[Debattista 2014]{debattista2014}).

Moreover, the velocity field for the young and old stellar populations are also found significantly different from the LAMOST and \emph{Gaia} common stars (\cite[Liu et al. 2017a]{liu2017a}). This may hint that the young and old populations have experienced different types of perturbation. 

\begin{wrapfigure}[21]{r}{0.4\textwidth}
\centering
\includegraphics[scale=0.4]{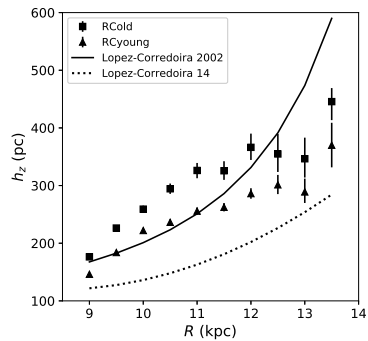}
\caption{The rectangles and triangles represent for the scale heights as functions of $R$ for the older and younger red clump stars (also see \cite[Wan et al. 2017]{wan2017}). The solid and dotted lines indicate the flaring scale heights from \cite[L\'opez-Corredoira et al. (2002)]{lopez2002} and \cite[L\'opez-Corredoira et al. (2014)]{lopez2014}.}\label{fig:3}	
\end{wrapfigure}

The previous observations of the Galactic stellar disk are mostly within the solar neighborhood or extend to larger distances along only very few lines of sight. Observations over a wider range of radius is requested so that the Milky Way can be mapped in much larger volume. The LAMOST spectroscopic survey (\cite[Cui et al. 2012]{cui2012}, \cite[Zhao et al. 2012]{zhao2012}), as one of the largest stellar spectroscopic surveys, has collected more than 8 million stellar spectra and has published more than 4 million of them in its third data release (DR3), recently. Within 40 kpc from the Sun, especially in the Galactic outer regions, the LAMOST data can provide sufficient samples for statistics (\cite[Deng et al. 2012]{deng2012}, \cite[Yao et al. 2012]{yao2012}). This enables us to give a panoramic view either in the spatial structural or in the kinematical features of the outer disk. In this talk, I highlight some new results on the star count and kinematics of the Galactic outer disk using the LAMOST giant stars.

\section{The disk structure from star counts}
\subsection{The size and shape of the outer disk}
Because LAMOST is a spectroscopic survey, estimation of the star count in 3 dimensional space using these data requires a careful selection correction. \cite[Liu et al. (2017b)]{liu2017b} corrected the post-observed selection effect by comparing the spectroscopic with photometric stellar density in color-magnitude diagram for each observation field. Then, they obtained the stellar density profile along the given line-of-sight. Their approach can be applied to any stellar sub-populations selected from metallicity, luminosity, age etc.

As a demonstration of the derived stellar density, \cite[Liu et al. (2017b)]{liu2017b} map the 2-D stellar density profile in $R$-$z$ plane with about 22000 carefully selected red giant branch (RGB) stars (\cite[Liu et al. 2014]{liu2014}, \cite[Wan et al. 2015]{wan2015} \cite[Tian et al. 2016]{tian2016}) from LAMOST DR3 data. Fig.~\ref{fig:1} shows the surface stellar density profile as a function of Galactocentric radius $R$ (hollow circles with error bars). It can be fitted with two components: an exponentially declining disk and a power-law stellar halo. We find that the disk is quite extended than previous works (e.g. \cite[Minniti et al. 2011]{minniti2011}). At $R=19$\,kpc, the fraction of the stellar halo (black solid line) is located at about 90\%, i.e. the disk component can still contributes to about 10\% at this radius. Moreover, neither truncation, nor bending-up nor -down feature is found in the outer disk.

The derived scale length of the exponential disk is $1.6\pm0.1$\,kpc, substantially smaller than the usually adopted value, which is $2.6\pm0.5$\,kpc (\cite[Bland-Hawthorn \& Gerhard 2016]{blandhawthorn2016}). The discrepancy may be due to that the scale length in the outer disk is not the same as that in the inner disk and solar neighborhood. Indeed, if we just consider the first three data points with $R<11$\,kpc, the derived scale length then becomes $2.37\pm0.02$\,kpc (see Fig.~\ref{fig:2}), which is approximately consistent with previous results. This may imply that the shape of the inner and outer disk is different. It mildly declines with $R$ when $R<11$\,kpc and then declines steeper. More detailed investigation on this point should be done in future.

\begin{figure}[htb]
\centering
	\includegraphics[scale=0.3]{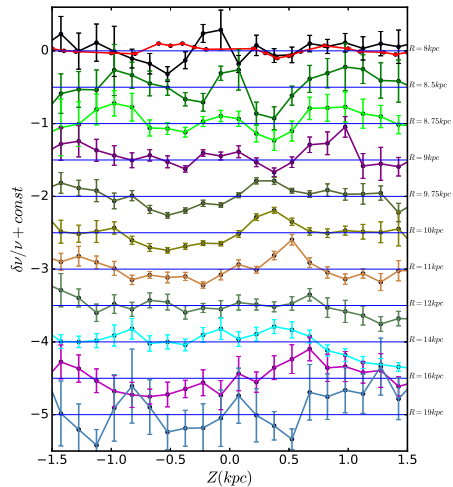}
	\caption{The residuals of the vertical density profiles after subtractiong the best fit models at various radii. From top to bottom the residuals are at $R=8$, 8.5, 8.75, 9, 9.75, 10, 11, 12, 13, 14, 16, and 19\,kpc.}\label{fig:4}
\end{figure}

\subsection{The flared disk}
With the selected red clump (RC) stars from the LAMOST DR3 data, we are able to investigate whether the flaring of the disk is related with the age of the stellar populations (\cite[Wan et al. 2017]{wan2017}). The younger RC stars (with mean age of about 2.7\,Gyr) and the older ones (with mean age of 4.6\,Gyr) show very similar flares, i.e. the scale height increases with $R$. Fig.~\ref{fig:3} shows that, although the overall values of the scale heights for the older population is slightly larger than those for the younger one, the slopes of the scale heights with respect to $R$ for the older and younger populations are $dh_z/dR=48\pm6$ and $40\pm4$\,pc\ kpc$^{-1}$, respectively. This means that no significant difference is found in the flares for the populations with ages between 2 and 5\,Gyr.

Meanwhile, from $R\sim8$ to $20$\,kpc, we can slice the RGB stars into various $R$ bins and derive the vertical density profile at each $R$ bin. Wang et al. in this volume shows that the scale heights of the thin disk measured in these $R$ bins increase with $R$ in similar way to the RC stars. Considering that the RGB stars are averagely older than the RC stars, this result again emphasizes  that the flaring disk may be similar in a quite large range of age.

\subsection{The wobbly disk}
\begin{wrapfigure}[19]{r}{0.4\textwidth}
	\centering
	\includegraphics[scale=0.3]{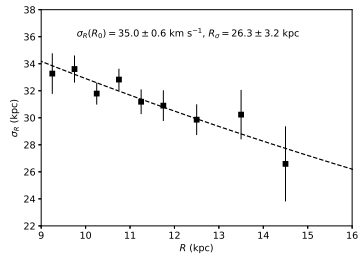}
	\caption{The radial profile (filled rectangles) of $\sigma_R$ derived from the RGB stars located at the Galactic anti-center direction. The dashed line indicates the best fit exponential model with the scale length of $R_\sigma=26.3\pm3.2$\,kpc.}\label{fig:5}
\end{wrapfigure}

Moreover, Fig.~\ref{fig:4} unveils multiple patterns of oscillations in a wide range of radius from the residuals of the vertical density profile for the RGB stars after the subtraction of the corresponding best-fit models.  At $R<9$\,kpc, the stellar densities at $z\sim0.5$\,kpc display dips, which are as much as 40\% lower than the best-fit ${\rm sech}^2$ vertical profiles. Between $9$ and $11$\,kpc, the north of the mid-plane has more stars and the residuals display a peak of about 20--40\%. At $12$--$14$\,kpc, the asymmetric stellar density turns around again, i.e. the south of the Galactic  mid-plane contains less star at $z>1.0$\,kpc. At around $R=19$\,kpc in which the Monoceros ring is supposed to be located (\cite[Newberg et al. 2002]{newberg2002}, \cite[Xu et al. 2015]{xu2015}), however, no significant over-density is found from the residual of the vertical stellar density profile.

\begin{figure}
	\centering
	\includegraphics[scale=0.5]{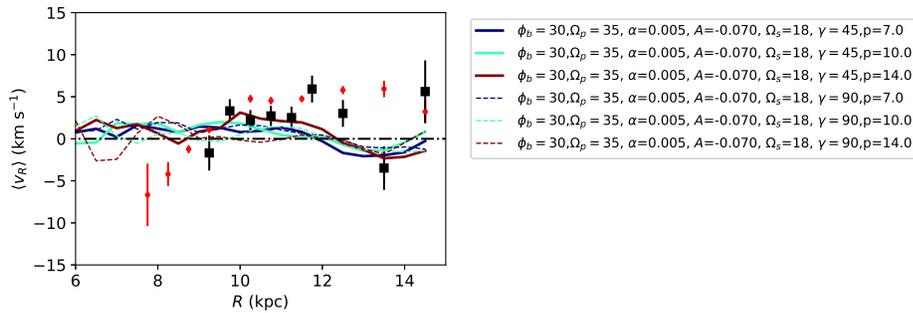}
	\caption{The black filled rectangles indicate the $\langle v_R\rangle$ derived from the RGB stars located in the Galactic anti-center direction. The red dots are the $\langle v_R\rangle$ of the older RC populations derived by \cite[Tian et al. (2016)]{tian2016}. The lines are the results of the test-particle simulations containing a rotating bar and 4 spiral arms with various parameters listed in the right side.}\label{fig:6}
\end{figure}

These results confirm the north near and south middle substructures argued by \cite[Xu et al. (2015)]{xu2015}. Moreover, at $R<9$\,kpc, the fact that the north contains less stars than the south is qualitatively consistent with \cite[Widrow et al. (2012)]{widrow2012}. 

Two possible reasons could explain the absent of the Monoceros ring. First, the previous discoveries of this substructure only used the stars above the Galactic mid-plane. The significant flare can produce an artificial over-density if the stars closed to the mid-plane are excluded in the analysis (\cite[Momany et al. 2006]{momany2006}). Second, if the substructure is real, i.e. not part of flared disk, then the stellar population should be fairly young so that it does not contain many giant stars and hence can not be effectively detected with RGB stars.

\begin{wrapfigure}[17]{r}{0.4\textwidth}
    \includegraphics[scale=0.4]{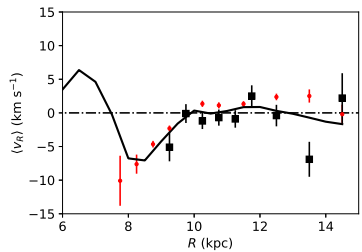}
	\caption{The black rectangles and red dots are same as in Fig.~\ref{fig:6} with $U_\odot=13$\,km\ s$^{-1}$ offset. The black line indicates the result of the test-particle simulation with pattern speed of 60 km\ s$^{-1}$\ kpc$^{-1}$.}\label{fig:7}
\end{wrapfigure}

\section{Stellar kinematics in outer disk}
\subsection{Velocity dispersion profile}\label{sec:sigprofile}
To derive the radial profile of the radial velocity dispersion, we select about 4000 RGB stars located within $175^\circ<l<185^\circ$ and $|z|<0.2$\,kpc. For these stars, their line-of-sight velocities are approximately equivalent to radial velocities. Fig.~\ref{fig:5} shows the radial profile of $\sigma_R$ (filled rectangles with error bars) and the best fit exponential model (dashed line). The best fit scale length of the radial profile is $R_\sigma=26.3\pm3.2$\,kpc, significantly larger than previous measurement (e.g. \cite[Lewis \& Freeman 1989]{lewis1989}). However, such a flat profile is quite consistent with the result of a torus model in the outer disk by \cite[Wang et al. (2017)]{wang2017}.

\subsection{Asymmetric motions}

\cite[Tian et al. (2016)]{tian2016} has made attempt to map the asymmetric radial motion using the younger and older RC stars and found that the radial velocity is negative between 8 and 9\,kpc and then becomes positive at $R>9$\,kpc.

With the RGB stars located in the Galactic anti-center direction (selection criteria are same as in section~\ref{sec:sigprofile}), we reproduce the similar asymmetric radial motion. Fig.~\ref{fig:6} shows the mean $v_R$ measured from the RGB stars as a function of $R$ with black rectangles. As a comparison, the radial motions derived from the older RC stars by \cite[Tian et  al. (2016)]{tian2016} are displayed with red dots. 

The asymmetric radial motion may be induced by the rotating bar, the local spiral arms, or the merging satellites. \cite[Grand et al. (2015)]{grand2015} argued that the bar may play an important role in the asymmetric motion in the outer disk. To investigate whether the bar is mainly responsible for the radial asymmetric motion, we run a set of 2-D test-particle simulations with an underlying axisymmetric logarithmic potential, a rotating bar (\cite[Dehnen 2000]{dehnen2000}), and a steady-state 4-arm logarithmic spiral potential. The simulations are implemented with \emph{galpy} (\cite[Bovy 2015]{bovy2015}). 

As shown in Fig.~\ref{fig:6}, the simulated radial velocity profiles with different configurations of the potential are displayed as lines with different colors. The parameters of the bar and the spiral structure is chosen as below. The angle between the major axis of the bar and the line connecting the Galactic center with the Sun is set at $\phi_b=30^\circ$, the pattern speed of the bar at $\Omega_p=35$\,km\ s$^{-1}$\ kpc$^{-1}$, and the relative strength of the bar at $\alpha=0.005$. The strength of the spiral structure is adopted as $A=0.07$ and the pattern speed of the spiral as $\Omega_s=18$\,km\ s$^{-1}$\ kpc$^{-1}$. The initial phase angle $\gamma$ is selected in between $45^\circ$ and $90^\circ$. And the pitch angle is chosen among $7^\circ$, $10^\circ$, and $14^\circ$. In general, the changing of the parameters of the bar can significantly change the radial velocity profile, especially the pattern speed. The simulations with the pattern speed of the bar at 35\,km\ s$^{-1}$\ kpc$^{-1}$ can roughly reproduce the observed radial velocity profile with $R>9$\,kpc, while it shows large inconsistency at $R<9$\,kpc. Note that the variations of the initial phase angle and the pitch angle of the spiral structure do not significantly alter the radial velocity profile.

Fig.~\ref{fig:7} shows another simulation with $\Omega_p=60$\,km\ s$^{-1}$\ kpc$^{-1}$. Meanwhile, the radial motion of the Sun is shifted from $U_\odot=9.58$\,km\ s$^{-1}$ to 13\,km\ s$^{-1}$. Then the simulated radial velocity profile can perfectly in agreement with the shifted observed result.

To summarize, it seems that either a slowly rotating bar with $\Omega_p=35$\,km\ s$^{-1}$\ kpc$^{-1}$ or a fast rotating bar with $\Omega_p=60$\,km\ s$^{-1}$\ kpc$^{-1}$ can fit the observed radial asymmetric motion. Even from the observations in the solar neighborhood, the two cases are not clearly discriminated from each other. For instance, \cite[P{\'e}rez-Villegas et al. (2017)]{perz2017} explained the Hercules stream with a slower pattern speed, while \cite[Dehnen (2000)]{dehnen2000} preferred  a faster pattern speed.

%
%

\end{document}